\documentclass[]{epl}
\usepackage{graphicx}
\usepackage{amsmath}
\usepackage{amssymb}

\newcommand{\bra}[1]{\langle #1 | \,}
\newcommand{\ket}[1]{\, | #1 \rangle}

\newcommand{\expv}[1]{\langle #1 \rangle}

\newcommand{\om}{\omega}
\newcommand{\Om}{\Omega}
\newcommand{\ga}{\gamma}

\newcommand{\de}{\delta}
\newcommand{\De}{\Delta}
\newcommand{\ka}{\kappa}
\newcommand{\sih}{\hat{\sigma}}
\newcommand{\Eh}{\hat{\cal E}}
\newcommand{\Psih}{\hat{\Psi}}
\newcommand{\psih}{\hat{\psi}}
\newcommand{\phih}{\hat{\phi}}
\newcommand{\Ih}{\hat{I}}
\newcommand{\Fh}{\hat{\cal F}}

\newcommand{\eps}{\epsilon}

\title{Giant nonlinearity and entanglement of single photons 
       in photonic bandgap structures}
\shorttitle{Entanglement of photons in photonic bandgap structures}

\author{I. Friedler\inst{1} \and G. Kurizki\inst{1} \and D. Petrosyan\inst{2}}
\shortauthor{I. Friedler \etal}

\institute{
 \inst{1}Department of Chemical Physics, Weizmann Institute of Science,
Rehovot 76100, Israel\\
 \inst{2}Institute of Electronic Structure \& Laser, FORTH, Heraklion 71110, 
Crete, Greece}


\begin{document}

\maketitle

\begin{abstract}
Giantly enhanced cross-phase modulation with suppressed spectral 
broadening is predicted between optically-induced dark-state polaritons 
whose propagation is strongly affected by photonic bandgaps of spatially
periodic media with multilevel dopants. This mechanism is shown to
be capable of fully entangling two single-photon pulses with high
fidelity.
\end{abstract}

\section{Introduction}
The main impediment towards the use of single photons in schemes for 
deterministic quantum logic and teleportation \cite{QCQI} as very robust 
and versatile carriers of quantum information \cite{linopt} is the weakness 
of optical nonlinearities in conventional media. A major trend aimed at the 
enhancement of optical nonlinearities exploits one-dimensional (1D-) periodic 
distributed Bragg reflectors and 2D- or 3D-periodic photonic crystals 
(PCs), where light can be slowed down or trapped via multiple reflections in 
the vicinity of photonic bandgaps (PBGs) \cite{pc-books}. Giantly enhanced 
nonlinearity has been predicted when dopants with transition frequencies 
within the PBG are implanted in the structure, so that light near such 
frequencies resonantly interacts with the dopants and is concurrently 
affected by the PBG dispersion \cite{Kuri2001}. The resulting soliton-like 
transmission of very weak pulses within the PBG while filtering out spurious 
noise is highly advantageous for classical communication \cite{josab}. 
However, this mechanism is incompatible with the goals of quantum logic 
and communications, particularly with photon-photon entanglement, because 
of the quantum noise associated with resonant field-atom interactions.

Another pathway to enhance the nonlinearities is based on electromagnetically 
induced transparency (EIT) in atomic media, which comes about when classical 
driving fields induce coherence between atomic levels and transform the 
weak fields into atom-dressed dark-state polaritons \cite{fllk,v0exp}. 
The ultrahigh sensitivity of the EIT polaritonic dispersion to a small 
field-induced Stark shift of its atomic level can result in an appreciable 
nonlinear phase shift, impressed by one ultraweak field upon another 
\cite{imam}. Notwithstanding this promising sensitivity, EIT-polariton 
entanglement by a large conditional phase shift of one photon in the 
presence of another (also known as cross-phase modulation) faces serious 
challenges in spatially uniform media. One drawback of these schemes has 
been the mismatch between the group velocities of the probe pulse moving 
as a slow EIT polariton and the nearly-free propagating signal pulse, 
which severely limits their effective interaction length and the maximal 
conditional phase-shift \cite{harhau}. To enable long interaction times and 
thus large conditional phase shifts in a medium of finite length (up to
a few centimeters), the group velocities of {\it both} interacting pulses 
should be small \cite{lukimam,petmal}. This, however, imposes a 
limitation on the photonic component of the signal polariton whose 
magnitude determines the phase shift. Copropagating pulses pose yet 
another difficulty: since the phase-shift of each pulse is proportional 
to the local intensity of the other pulse, different parts of the 
interacting pulses acquire different phase shifts, which results in 
their spectral broadening.

Here we put forward a mechanism that may produce strong photon-photon 
interactions along with suppressed quantum noise and give rise to their 
entanglement with high fidelity, by combining the advantages of 
their dispersion in PBG structures and of the strongly enhanced nonlinear 
optical coupling achievable via EIT in an appropriately doped medium.
The main idea is that a single-photon signal pulse is adiabatically 
converted into a standing-wave polaritonic excitation inside the periodic 
structure. This trapped polariton, having an appreciable photonic component, 
can impress a large, spatially-uniform phase shift upon a slowly 
propagating probe polariton. This task can further be facilitated
by employing 2D- or 3D-periodic structures with defects where the two 
pulses interact via tightly confined modes \cite{pc-books,kivshar,noda}.

\section{Photon-photon interaction in PBG structures}
\begin{figure}[t]
\centerline{\includegraphics[width=14.0cm]{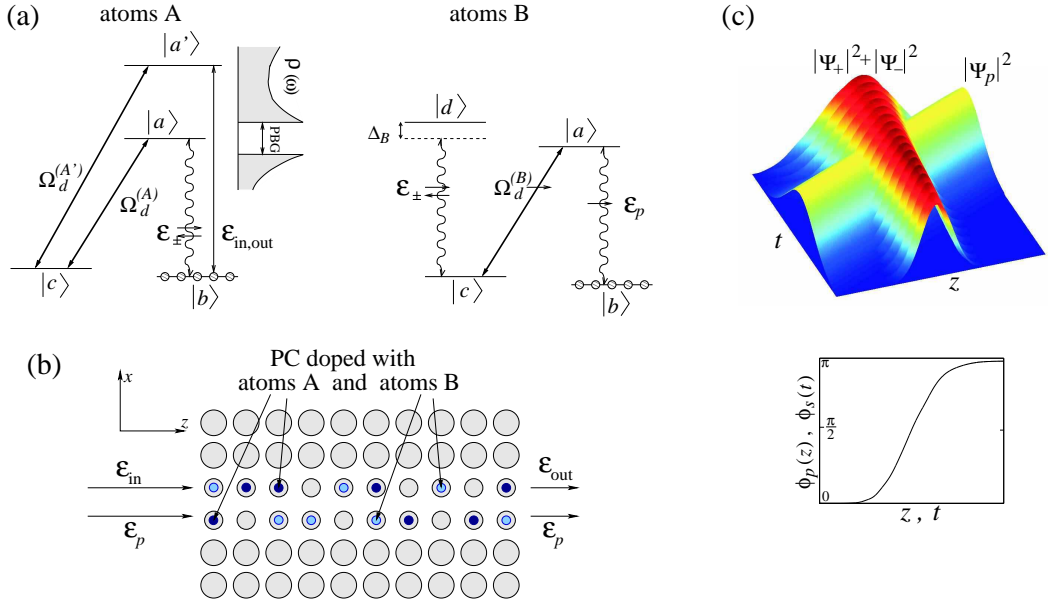}} 
\caption{(a)~Level scheme of atomic species A and B. Atoms A convert the 
input signal field ${\cal E}_{\rm in}$ at frequency outside the PBG to the 
trapped field ${\cal E}_{\pm}$ at frequency inside the PBG. 
Atoms B provide EIT for the probe field ${\cal E}_p$ and its 
cross-coupling with the signal field ${\cal E}_{\pm}$.
(b) 2D-periodic photonic crystal (PC) having the density of modes 
$\rho(\om)$, is doped with atomic species A and B.
(c)~Interaction and the resulting phase-shift of the probe and signal pulses.} 
\label{as-pr}
\end{figure}

The proposed scheme is based upon a periodic structure containing uniformly 
distributed dopants -- atoms A and B (see fig. \ref{as-pr}). Atoms A, 
having double-$\Lambda$ level configuration and interacting with 
classical driving fields on the transitions 
$\ket{c}_{A} \to \ket{a}_{A},\ket{a^{\prime}}_{A},$ with the Rabi 
frequencies $\Om^{(A,A^{\prime})}_d$, respectively, facilitate the 
trapping of the signal pulse inside the periodic structure, 
by converting its frequency from outside to inside of the PBG. On the 
other hand, atoms B, having N level configuration and interacting with
the $\Om^{(B)}_d$ driving field on the transition 
$\ket{c}_{B} \to \ket{a}_{B}$, serve to simultaneously slowdown 
the probe pulse and cross-couple it with the signal. 

The following procedure is foreseen to this end. Initially, all  
atoms A are in the ground state $\ket{b}_A$, the driving fields 
$\Om_d^{(A)} = 0$ and $\Om_d^{(A^{\prime})} > T_{\rm in}^{-1}$, 
where $T_{\rm in}$ is the temporal width of the input signal pulse 
${\cal E}_{\rm in}$. The carrier frequency of ${\cal E}_{\rm in}$ is 
outside the PBG, close to the $\ket{b}_A \to \ket{a^{\prime}}_A$ 
transition frequency, so that the usual EIT for the input signal 
due to the $\Lambda$ configuration 
$\ket{b}_A \leftrightarrow \ket{a^{\prime} }_A \leftrightarrow \ket{c}_A$
is realized. Upon entering the medium, the signal pulse is slowed down 
and spatially compressed, by a factor of $v_s^{\prime}/c \ll 1$, to 
the length $z_{\rm loc} \simeq  T_{\rm in} v_s^{\prime}$, where 
$v_s^{\prime} \propto |\Om_d^{(A^{\prime})}|^2$ is its group velocity 
inside the medium. Once the signal pulse has fully accommodated in the 
medium of length $L$, which requires that $z_{\rm loc} < L$, the driving 
field $\Om_d^{(A^{\prime})}$ is adiabatically switched off. As a result, 
the signal is stopped, its photonic component being converted into 
the stationary atomic (Raman) coherence $\sigma^{(A)}_{bc}$ \cite{fllk,v0exp}. 
Next, the driving field $\Om_d^{(A)}$ is adiabatically switched on
to a value $\Om_d^{(A)} > \Om_d^{(A^{\prime})}$, converting the atomic 
coherence into the signal field ${\cal E}_{\pm}$, whose frequency is 
inside of the PBG and amplitude is larger than that of the input signal
${\cal E}_{\rm in}$ by a factor of $\sqrt{v_s/v_s^{\prime}} \simeq 
\Om_d^{(A)} / \Om_d^{(A^{\prime})} >1$ \cite{v0exp}. Due to the Bragg 
scattering of the forward and backward propagating components of the 
signal pulse with the wave vectors $\pm k$, it remains localized (trapped) 
within the medium \cite{pc-books}. Both components ${\cal E}_{\pm}$ of 
the signal field dispersively interact with atoms B via transition 
$\ket{c}_{B} \to \ket{d}_{B}$ with the detuning $\De_B$. This off-resonant 
interaction causes an ac Stark shift of level $\ket{c}_B$, thereby strongly 
affecting the EIT dispersion for the probe field ${\cal E}_p$, which 
interacts with atoms B on the transition $\ket{b}_B \to \ket{a}_B$ 
\cite{imam}. For a large enough product of the signal-field intensity 
$|{\cal E}_{\pm}|^2$ and interaction time $L/v_p$ ($v_p$ being the probe 
group velocity), both pulses accumulate a {\it uniform} conditional 
phase-shift which can reach $\pi$ (see fig. \ref{as-pr}(c)). Finally, 
reversing the sequence that resulted in trapping of the signal pulse, 
its frequency is converted back to the original frequency and the 
${\cal E}_{\rm out}$ pulse leaves the medium.

Let us now consider the scheme more quantitatively. To describe
the quantum properties of the medium, we use collective slowly
varying atomic operators 
$\sih^{(\iota)}_{\mu \nu}(z,t) = \frac{1}{N_{\iota}^z} 
\sum_{j=1}^{N_{\iota}^z} \ket{\mu_j}_{\iota \iota}
\bra{\nu_j} e^{-i \om_{\mu \nu}^{(\iota)}t}$, 
averaged over a small but macroscopic volume containing many dopants of
species $\iota=A,B$ around position $z$ \cite{fllk}: $N_{A,B}^z =
(N_{A,B}/L) \, \upd z \gg 1$, where $N_{A,B}$ is the total number of the
corresponding dopants. The quantum radiation is described by the 
traveling-wave (multimode) electric field operators  
$\Eh_{\pm}(z,t) = \sum_{q} a_{\pm}^{q}(t)e^{\pm iq z}$ and 
$\Eh_p (z,t) = \sum_{q} a_{p}^{q}(t) e^{i q z}$, where $a_j^q$ are the 
annihilation operators for the field mode with the wave-vector $k_j+q$, 
$k_j$ being the carrier wave vector of the corresponding field. 
These single-mode operators possess the 
standard bosonic commutation relations 
$[a_{i}^{q},a_{j}^{q^{\prime}\dagger}]=\de_{ij} \de_{qq^{\prime}}$, 
which yield $[\Eh_{i}(z),\Eh_{j}^{\dagger}(z^{\prime})]
= L \de_{ij} \de(z-z^{\prime})$.

Using the standard technique \cite{fllk}, we perturbatively solve the
Heisenberg equations for the atomic coherences $\sih^{(\iota)}_{\mu \nu}$
under EIT conditions \cite{fllk,v0exp,imam,harhau,lukimam,petmal} and 
substitute the solution into the propagation equations for the slowly 
varying field operators $\Eh_j (z,t)$. These operators are related to 
the polariton operators $\Psih_{\pm} = \Eh_{\pm}/\cos \theta_A$ and 
$\Psih_p = \Eh_p/\cos \theta_B$, which represent the coupled excitation 
of the corresponding field and atomic Raman coherence  $\sih^{(\iota)}_{bc}$.
The mixing angles $\theta_{A,B}$ are defined via 
$\tan \theta_{A,B} = g_{A,B} \sqrt{N_{A,B}}/\Om_d^{(A,B)}$,
where $g_{A,B} = \wp_{ab}^{(A,B)} \sqrt{ \om_{ab}^{(A,B)}/(2 \hbar \eps_0 SL)}$
are the atom-field coupling constants, $\wp_{\mu \nu}^{(\iota)}$ being 
the corresponding atomic dipole matrix element and $S$ the cross-sectional 
area of the quantum fields. Note that the amplitude of the photonic component
of each polariton is proportional to the $\cos \theta_{\iota}$ of the 
corresponding mixing angle $\theta_{\iota}$. Under the Bragg resonance 
condition $k \simeq \pi/p_s$, where $p_s$ is the structure period, the 
equations of motion for polaritons are obtained as
\begin{subequations}
\label{pols}
\begin{eqnarray}
\left(\frac{\partial}{\partial t}\pm v_s
\frac{\partial}{\partial z}\right)\Psih_{\pm} &=&  -\ka_s \Psih_{\pm}
+ i \eta \Ih_p \Psih_{\pm} + i \beta \Psih_{\mp} + \Fh_s , \label{cpol}  \\ 
\left(\frac{\partial}{\partial t}+ v_p
\frac{\partial}{\partial z}\right)\Psih_p &=& -\ka_p \Psih_p 
+ i \eta \Ih_s \Psih_p + \Fh_p. \label{ppol} 
\end{eqnarray}
\end{subequations}
Here $v_s = c \cos^2\theta_A$ and $v_p = c \cos^2\theta_B$ 
are the group velocities, $\Ih_s \equiv \Psih_+^{\dagger}\Psih_+ 
+ \Psih_-^{\dagger}\Psih_-$ and ${\Ih_p \equiv \Psih_p^{\dagger}\Psih_p}$ 
are the intensities (excitation numbers) of the signal and probe polaritons, 
respectively;  $\ka_{s,p} = \ga_{bc}^{(A,B)} \sin^2 \theta_{A,B}$ are
the absorption rates, $\ga_{bc}^{(\iota)}$ being the Raman coherence decay 
rate of the corresponding atoms, $\Fh_{s,p}$ are the $\de$-correlated 
noise operators associated with the relaxation; 
$\eta = [1+i \ga_d^{(B)}/(2 \De_B)]\cos^2\theta_A \sin^2\theta_B 
g_B^{\prime 2} /\De_B$ is the cross-coupling rate between the polaritons, 
$g_B^{\prime} = \wp_{dc}^{(B)} \sqrt{ \om_{dc}^{(B)}/(2 \hbar \eps_0 SL)}$
being the atom-field coupling on the transition $\ket{c}_B \to \ket{d}_B$
and $\ga_d^{(B)}$ the decay rate of $\ket{d}_B$ (for $\De_B \gg \ga_d^{(B)}$, 
the cross-absorption vanishes \cite{imam,harhau,lukimam} and then $\eta$, 
being purely real, represents the cross-phase modulation rate).
Finally, $\beta = \frac{1}{2} \de \om_{\rm PBG} \cos^2 \theta_A$ is the Bragg 
reflection rate, $\de \om_{\rm PBG}$ being the PBG bandwidth.

Equations (\ref{pols}) constitute the starting point of our analysis. 
Their general solution, for arbitrary initial/boundary conditions of 
the traveling-wave quantized polaritons $\Psih_{p,\pm}$ is not known. 
When absorption is negligible (see below), for a given time- 
and space-dependence of the signal-polariton intensity $\Ih_s(z,t)$, 
the solution for the probe is
\[
\Psih_p(z,t) = \Psih_p(0,\tau) \exp \left[  i \frac{\eta}{v_p} 
\int_0^z \Ih_s(z^{\prime},\tau+z^{\prime}/v_p) \, \upd z^{\prime} \right],
\]
where $\tau=t-z/v_p$ is the retarded time.
An analytic solution for the two counter-propagating components of
the signal polariton can be obtained in the long probe limit, i.e., 
when the spatial dependence of the probe-polariton intensity is 
negligible on the scale of $z_{\rm loc}$: $\Ih_p(z,t) \simeq \Ih_p(t)$. 
This requires that $v_p T_p > z_{\rm loc}$, where $T_p$ is the duration 
of the probe pulse (its spectral width being 
$\de \om_p \sim T_p^{-1} < v_p/z_{\rm loc}$). 
Then eq.~(\ref{cpol}) is soluble by the Fourier transform
$\Psih_{\pm}(z,t) = \frac{1}{2 \pi} \int \upd q \,
e^{\pm i q z} \psih_{\pm}(q,t)$, with the result
\begin{subequations}
\label{spolms}
\begin{eqnarray}
\psih_{+}(q,t)&=&\psih_{+}(q,0)\: e^{ i \phih_s(t)} 
\left[ \cos(\chi t) - i \frac{q v_s}{\chi} \sin(\chi t) \right] , \\
\psih_{-}(q,t) &=& i \psih_{+}(-q,0) \: e^{i \phih_s(t)} 
\frac{\beta}{\chi} \sin(\chi t),
\end{eqnarray}
\end{subequations}
where $\chi=\sqrt{q^2 v_s^2+\beta^2}$ and 
$\phih_s(t)=\eta \int_0^t \Ih_p(t^{\prime}) \, \upd t^{\prime}$, with
$\Ih_p(t)= \Psih_p^{\dagger}(0,\tau) \Psih_p(0,\tau)$. Thus all the spatial 
modes $\psih_{\pm}(q,t) = \int \upd z \, e^{\mp i q z} \Psih_{\pm}(z,t)$ 
of the signal polariton acquire the same $q$-independent phase shift 
$\phih_s(t)$. It follows from eqs. (\ref{spolms}) that a signal polariton 
composed of modes with $|q| \ll \beta/v_s$ will be strongly trapped 
inside the medium, its wavepacket periodically cycling between the forward
and backward components while interacting with the probe
polariton, yielding
\begin{subequations}
\label{polsol}
\begin{eqnarray}
\Psih_{+}(z,t)&=& \Psih_{+}(z,0) \: e^{i \phih_s(t)} \cos(\beta t) ,\\
\Psih_{-}(z,t)&=& i \Psih_{+}(z,0) \: e^{i \phih_s(t)} \sin(\beta t) ,\\
\Psih_p(z,t)&=& \Psih_p(0,\tau) \: e^{i \phih_p(z)} ,
\end{eqnarray}
\end{subequations}
where $\phih_p(z) = \frac{\eta}{v_p} 
\int_0^z \Ih_s(z^{\prime}) \, \upd z^{\prime}$,
with $\Ih_s(z) = \Psi_+^{\dagger}(z,0) \Psi_+(z,0)$, is the probe phase-shift.

Let us dwell upon the approximations involved in the derivation of 
eqs. (\ref{polsol}). During the conversion of the signal pulse into 
a standing-wave polaritonic excitation inside the periodic structure,
the nonadiabatic corrections resulting in its dissipation are negligible 
provided the medium is optically thick, $\varsigma_{A} \rho_A L \gg 1$, 
where $\varsigma_{A}$ is the resonant absorption cross-section for the 
transition $\ket{b}_A \to \ket{a^{\prime}}_A$ and $\rho_A=N_A/(SL)$ is the 
density of atoms $A$ \cite{fllk,v0exp}. After the signal pulse has been 
trapped in the PBG, due to nonzero values of $q$, its gets spatially 
distorted (spreads) at a rate 
$\ka_d \simeq q^2 v_s^2 /(\pi\beta)$ ($0 \leq |q|< \beta/v_s$). 
We can estimate the bandwidth of the signal pulse from its spatial extent
as $\de q \sim v_s/(c L) <|\Om_d^{(A)}|^2/(\ga_a^{(A)} c)$, 
thus obtaining the upper limit for the distortion rate 
$\ka_d \leq 2 v_s^3 /(\pi c  L^2 \de \om_{\rm PBG})$. On the other hand, 
the bandwidth of the probe is limited by the length of the medium via 
$\de \om_p < |\Om_d^{(B)}|^2 [g_B^2 N_B \ga_a^{(B)}L/c]^{-1/2} 
= |\Om_d^{(B)}|^2 k_p [\ga_a^{(B)} \sqrt{3 \pi/2 \rho_B L}]^{-1}$, 
where $\rho_B=N_B/(SL)$ is the density of atoms $B$ \cite{scully}.
Finally, the interaction time $t_{\rm int}= L/v_p$ is limited by 
$t_{\rm int} \times \max\{\ka_d, \ka_s, \ka_p \} \ll 1$, and so
the fidelity of the cross-phase modulation is given by 
$F=\exp[-(\ka_d + \ka_s+\ka_p)L/v_p]$. 

Consider first the classical limit of eqs. (\ref{polsol}), where the 
operators $\Psih_{p,\pm}$ and $\Ih_{p,s}$ are replaced by c-numbers. 
Then for two single-photon pulses 
$\frac{1}{L}\int I_s \, \upd z = \frac{v_p}{L} \int I_p \, \upd t \simeq 1$,
the conditional phase shift accumulated by the probe and signal fields 
during the interaction is given by 
\begin{equation}
\phi_p = \phi_s  = \frac{g_B^{\prime 2} L \cos^2\theta_A \tan^2\theta_B} 
{c \De_B}  \equiv \phi. \label{phsh_cls}
\end{equation}
Note that the phase shift is proportional to the intensity of the photonic 
component of the signal polariton, as attested  by the presence 
of the $\cos^2\theta_A$ term in the nominator of eq. (\ref{phsh_cls}).
For realistic experimental parameters, relevant to a doped periodic 
structure discussed below, one can obtain $\phi \simeq \pi$ 
(see fig. \ref{as-pr}(c)) with the fidelity $F \simeq 1$.

We now turn to the fully quantum treatment of the system. To compare the
classical and quantum pictures, we consider first the evolution of input 
wavepackets composed of the multimode coherent states 
$\ket{\alpha_p} \otimes \ket{\alpha_+} \otimes \ket{0_-}= 
\prod_{q_p} \ket {\alpha_p^{q_p}} \otimes \prod_{q} \ket {\alpha_+^{q}}
\otimes \prod_{q^{\prime}}  \ket {0_-^{q^{\prime}}}$. States $\ket{\alpha_p}$
and $\ket{\alpha_+}$ are the eigenstates of the input fields operators
$\Eh_p (0,t)$ and $\Eh_+ (z,0)$ with eigenvalues
$\alpha_p(t) = \sum_{q_p} \alpha_p^{q_p} e^{-i q_p c t}$
and $\alpha_+(z) = \sum_{q} \alpha_+^{q} e^{i q z}$, respectively. 
The expectation values for the fields are then obtained as
\begin{subequations}
\label{expval}
\begin{eqnarray}
\expv{\Eh_p (z,t)} &=& \alpha_p(\tau)  
\exp \left[ \frac{e^{i \phi c/v_s}-1}{L} 
\int_0^z |\alpha_+(z^{\prime})|^2 \, \upd z^{\prime} \right] \, , \\ 
\expv{\Eh_{\pm} (z,t)} &=& \alpha_{+} (z)  
\exp \left[ \frac{e^{i \phi}-1}{L} c
\int_0^t |\alpha_p(\tau^{\prime})|^2 \, \upd t^{\prime} \right] \times 
\left[
\begin{array}{c} \cos(\beta t) \\ i \sin(\beta t) \end{array} 
\right] \, .
\end{eqnarray}
\end{subequations}
These equations are notably different from those obtained for single-mode 
\cite{sami} and multimode copropagating fields \cite{lukimam,petmal} 
because all parts of the probe pulse interact with the whole signal 
pulse (and vice versa), as is manifest in the space (time) integration. 
Similarly to the cases discussed in \cite{lukimam,petmal,sami}, 
eqs. (\ref{expval}) reproduce the classical result only 
in the limit $\phi c/v_s \ll 1$, yielding 
$\phi_p = \frac{\eta c}{v_p v_s} \int |\alpha_+|^2 \, \upd z^{\prime}$ 
and $\phi_s = \frac{\eta c}{v_p} \int |\alpha_p|^2 \, \upd t^{\prime}$. 
A $\pi$ phase shift is then obtained for 
$\frac{c}{L v_s} \int |\alpha_+|^2 \, \upd z^{\prime} = 
\frac{c}{L} \int |\alpha_p|^2 \, \upd t^{\prime} = \pi /\phi$.  
This restriction on the classical correspondence of coherent states 
comes about since, for large enough cross-phase modulation rates $\eta$, 
these states exhibit periodic collapses and revivals as $\phi$ changes 
from 0 to $2 \pi$, which limits their usefulness for quantum information 
applications. 

Consider finally the input state $\ket{\Phi_{\rm in}} =
\ket{1_{p}} \otimes \ket{1_{+}} \otimes \ket{0_-}$, consisting of
two single-excitation polariton wavepackets $\ket{1_{p,+}} = \sum_{q}
\xi_{p,+}^q \ket{1_{p,+}^{q}}$, where the Fourier
amplitudes, normalized as $\sum_{q} |\xi_{p,+}^q|^2 =1$, define
the spatial envelopes $f_{p,+}(z) = \sum_{q} \xi_{p,+}^q e^{i q z}$ 
of the probe and forward signal pulses that initially (at $t=0$) are 
localized around $z=0$ and $z=L/2$, respectively (fig. \ref{as-pr}(c)).
During the evolution, the state of the system evolves according to 
$\ket{\Phi(t)} = \exp \left(-\frac{i}{\hbar} 
\int_0^t H_{\rm int} \, \upd t^{\prime} \right) \ket{\Phi_{\rm in}}$, where 
$ H_{\rm int} = -\frac{\hbar}{L}\int \upd z \,  
[\eta \Psih_p^{\dagger}\Psih_p(\Psih_+^{\dagger}\Psih_+ + 
\Psih_-^{\dagger}\Psih_-) 
+ \beta (\Psih_+^{\dagger} \Psih_ - + \Psih_-^{\dagger}\Psih_+)]$
is the effective interaction Hamiltonian whose first and second terms 
commute. The implicit time-dependence of the effective Hamiltonian, due
to the propagation of the probe-polariton pulse with the group velocity
$v_p$, is contained in the operator $\Psih_p = \Psih_p(z-\zeta)$, with
$\zeta = v_p t$. After the interaction, at time $t_{\rm out} > L/v_p$, 
the output state of the system is 
\begin{equation}
\ket{\Phi_{\rm out}} = e^{i \eta L/v_p} \ket{1_{p}}\otimes
[ \cos(\beta t_{\rm out}) \ket{1_{+}} \otimes \ket{0_-} 
+ i \sin(\beta t_{\rm out}) \ket{0_{+}} \otimes \ket{1_-}] , \label{sph}
\end{equation}
where $\ket{1_-} \equiv \sum_{q} \xi_{+}^q \ket{1_-^{-q}}$. Thus, while the 
signal pulse periodically cycles between the forward and backward modes, the 
combined state of the system acquires an overall conditional phase shift 
$\phi = \eta L/v_p$. When $\phi = \pi$, transformation (\ref{sph}) corresponds 
to the truth table of the {\it universal} controlled-phase (\textsc{cphase}) 
logic gate between the two photons representing qubits, which can be used to 
realize arbitrary unitary transformation \cite{QCQI}. 

\section{Possible experimental realizations}
An $x-z$ 2D-periodic lattice of dielectric rods or semiconductor stacks 
(fig. \ref{as-pr}(b)) \cite{pc-books,kivshar,noda}, with controlled structural 
defects and mirror confinement in $y$, appears to be the most suitable 
structure for realizing polaritonic entanglement of two single-photons, 
since both the signal and the probe pulses may be confined in the $x-y$ 
directions in the vicinity of a ``defect'' row forming a PC waveguide, 
thereby avoiding diffraction losses and focusing the fields to a radius 
of $\sim 1\:\mu$m \cite{pc-books,kivshar}. Using the double-$\Lambda$ dopants
(atoms A), the signal pulse can be trapped in the PC, with the localization 
distance $z_{\rm loc}$ extending over many periods, and interact with
the probe pulse via atoms B. Expressing the atom-field coupling constants 
$g$ through the decay rate $\ga$ of the corresponding excited state as 
$ g = 3 \pi c \ga/(2 k^2 SL)$, and assuming that 
$\ga_a^{(A)} \simeq \ga_a^{(B)}$ and 
$g_A^2 N_A \gg |\Om_d^{(A)}|^2$ ($v_s \ll c$), we have from 
eq. (\ref{phsh_cls}) $\phi \simeq 3 \pi \ga_d \, (2 k^2_p \De_B S)^{-1}   
(\Om_d^{(A)}/\Om_d^{(B)})^2 (\rho_B/\rho_A)$.
Among the possible dopants, III-V or II-VI $n$-doped semiconductor quantum 
dots, having large dipole moments and level structure conducive to EIT 
\cite{qdopt}, could be the best choice for our scheme, provided high 
densities can be achieved. In such single-electron doped QDs, the spin 
degeneracy of the ground and the lowest and higher excited (charged exciton 
or trion) states can be lifted with a magnetic field \cite{qdopt}, 
realizing the level scheme of fig. \ref{as-pr}(a), where atoms A and B can 
spectroscopically be selected (via optical pumping or spectral hole burning) 
from the inhomogeneous ensemble of QDs. Thus states $\ket{b}$ and $\ket{c}$ 
are represented by the Zeeman-split spin-up and spin-down states of the 
conduction-band electron in the QD. The excited states $\ket{a}_A$ and 
$\ket{a'}_A$ of atoms A can be the first and the second (or higher) exciton 
states, while states $\ket{a}_B$ and $\ket{d}_B$ of atoms B are the Zeeman 
sublevels of the lowest excitonic state. Assuming the parameters
$L \simeq 0.1\:$cm, $S \simeq 10^{-8}\:$cm$^{2}$,  
$\de \om_{\rm PBG} \sim 10^{14}\:$rad/s \cite{pc-books,kivshar},
$\rho_{A,B} \simeq 10^{12}\:$cm$^{-3}$,
$\om_p \simeq 3 \times 10^{15}\:$rad/s, 
$\Om_d^{(A)} \simeq 5 \times 10^8\:$rad/s, 
$\Om_d^{(B)} \simeq 2 \times 10^7\:$rad/s, 
$\De_B \simeq 10^8\:$rad/s,
$\ga_{a,d} \simeq 10^{7}\:$s$^{-1}$, and 
$\ga_{bc} \simeq 10^{4}\:$s$^{-1}$ \cite{qdopt}, we obtain $\phi \simeq \pi$ 
with the fidelity $F \geq 0.98$\%, the main limiting factor being the decay 
of Raman coherence $\ga_{bc}$. Other contenders for observing the proposed
effects include periodic structures fabricated from rare-earth doped 
crystals, such as Pr:YSP, in which high-fidelity EIT has experimentally 
been demonstrated \cite{EIT_solid}, or cryogenically cooled diamond with
high density of nitrogen-vacancy defect centers \cite{N-V}.

\section{Conclusions}
To summarize, we have proposed a new class of multimode quantum-field 
interactions involving quantized EIT-polaritons in PBG structures. 
We have shown that such interactions allow efficient cross-phase modulation 
between a propagating probe pulse and a trapped signal pulse, whose 
localization is achieved by an adiabatic four-wave mixing process that
pulls its frequency into the PBG. This localization allows multiply 
repeated interaction of the signal with the entire probe pulse. As a 
result, the combined two-photon state of the system can acquire a 
conditional $\pi$ phase-shift, which corresponds to the universal
\textsc{cphase} logic gate. The phase shift is spatially-uniform and 
the process may have high fidelity. The experimental realization of 
the predicted effects requires the fabrication of periodic structures
with large densities of optically active dopants \cite{qdopt,EIT_solid,N-V}, 
which may also find useful applications in laser technology, optical 
communication or quantum computation. We note that a similar regime of 
giant cross-phase modulation with suppressed spectral broadening is also 
realizable in cold atomic vapors using optically-induced PBGs \cite{opt-pbg},
on which we intend to report elsewhere. The proposed scheme 
may pave the way to quantum information applications such as deterministic 
all-optical quantum computation, dense coding and teleportation \cite{QCQI}.

\vspace{-0.3cm}
\acknowledgments
\vspace{-0.3cm}
We gratefully acknowledge stimulating discussions with A. Andr\'{e}
and M. D. Lukin.
This work was supported by the EC (QUACS, ATESIT and PHOREMOST Networks),
ISF and Minerva.

\end{document}